\documentclass[aps,pre,amssymb,preprint]{revtex4-1}
\usepackage{graphicx}
\usepackage{hyperref}
\usepackage{xr} 
\usepackage{url}
\usepackage{amsmath}
\usepackage{setspace}
\usepackage{xcolor}

\begin{document}
\title{Granular bed consolidation, creep and armoring under subcritical fluid flow}
\author{Benjamin Allen and Arshad Kudrolli}
\affiliation{Department of Physics, Clark University, Worcester, MA 01610}

\date{\today}

\begin{abstract}
We show that a freshly sedimented granular bed settles and creeps forward over extended periods of time under an applied hydrodynamic shear stress 
which is below the critical value for bedload transport.  The rearrangements are found to last { over 
a time scale which is millions of times} the sedimentation time scale of a grain in the fluid. Compaction occurs uniformly throughout the bed, but creep is observed to decay exponentially with depth, and decreases over time. The granular volume fraction in the bed is found to increase logarithmically, saturating at the random close packing value $\phi_{rcp} \approx 0.64$, while the surface roughness is observed to remain essentially unchanged. 
{ We demonstrate that an increasingly higher shear stress is required to erode the bed after a sub-critical shear is applied which results in an increase in its volume fraction.}  Thus, we find that bed armoring occurs due to a deep shear-induced relaxation of the bed towards the volume fraction associated with the glass transition. 
\end{abstract}

\maketitle
\section{Introduction}
The response of a granular bed to a fluid flowing over its surface is important to the evolution of rivers and beaches, among many other geophysical and industrial systems~\cite{lobkovsky07,berhanu12}. The onset of erosion and bed transport has been studied in the field and in large scale flumes for over a century~\cite{shields36}. { But, the results have been mixed because of the difficulty in making measurements and the complexity of the system~\cite{buffington97,recking12}.} Therefore, a number of experiments and simulations have been performed more recently focusing on idealized monodisperse granular beds under model fluid flows~\cite{charru04,mouilleron09,derksen11,houssais15,allen17}. It has been shown that the condition to dislodge a grain on a rough bed corresponds to the torque balance condition, and a material dependent formula for the onset of a grain's motion has been developed~\cite{kudrolli16}.  The grain dynamics above the onset of  bedload transport have been investigated with  index matching techniques~\cite{tsai03,lobkovsky08,houssais15,allen17,mouilleron09}, and granular velocities have been shown to decay exponentially into the bed under steady state conditions~\cite{allen17}. 
 Nonetheless, many important questions pertaining to the onset of erosion and the state of the bed remain. 

{ The dimensionless hydrodynamic shear stress $\tau_c$ 
given by the ratio of the force acting on a grain tangential to the bed surface and its buoyancy subtracted weight is typically used to understand the driving conditions at the bed surface. Then, the corresponding critical hydrodynamic shear stress $\tau^*_c$, when sustained grain motion occurs at the bed surface, is often used to estimate erosion thresholds~\cite{shields36,buffington97}. 
Typically it is assumed that the shear stress is determined under steady state conditions and thus $\tau^*_c$ can be expected to be constant and history independent.  

However, a number of studies on beds composed of monodisperse beads have noted transient bed motion under prescribed steady state driving conditions~\cite{charru04,houssais15,ouriemi09,hong15,allen17}. 
Because the driving and thus $\tau^*$ is assumed to be unchanged at the bed surface, it is implied that the bed  becomes resistant to erosion up to a higher $\tau^*$ as the transients die out. Thus, such a granular bed which becomes resistant to erosion over time was noted to armor 
by Charru, et al~\cite{charru04}.  It may be noted that bed armoring, also sometimes called paving, is commonly known to occur in polydisperse beds due to systematic erosion of smaller particles which leave behind larger or heavier particles that require higher shear stress to erode~\cite{parker82}. That mechanism is absent in monodisperse beds and thus cannot be the explanation for the observed armoring in the newer model experiments~\cite{charru04,ouriemi09,hong15}.   
These reports have typically dealt with these observed time-dependence by using variable wait times to obtain values observed after the fast transients have subsided. To our knowledge, few systematic studies have been conducted to check if a true asymptotic time-independent regime is reached or for that matter any systematic study on the effect of varying the waiting time on the observed $\tau^*$ required to erode particles. 
These same studies also report a small decrease in bed height. The small decrease by itself is not sufficent to explain the relatively large magnitude of the noted armoring in those studies.  Thus, it remains unclear if the armoring arises due to surface rearrangements, whereby grains which protrude further into the fluid are selectively dislodged~\cite{fenton77}, or if the resistance to shear arises due to rearrangements deeper within the bed and any resulting changes to the fluid flow profile inside and near the bed surface.

Further, in the case of dry frictionless granular materials subject to linear boundary driven shear, it has been noted that grains rearrange even under vanishingly small applied stress~\cite{clark15}. This can lead to creep like behavior in granular materials subject to shear and has been well studied in case of dry granular materials~\cite{komatsu01}. Indeed, in geophysics, long-term stress relaxation is known to occur post-slip across fault lines~\cite{smithwyss68,bucknam78,marone91,nasuno97}. In other dry granular systems in gravity, vibration induced rearrangements of granular materials is observed to lead to slow compaction corresponding to logarithmic increase in granular volume fraction~\cite{knight95}.  These systems have been compared to out-of-equilibrium glassy dynamics with the logarithmic behavior that comes from a combination of wide range of size and time scales over which rearrangements take place~\cite{ribiere05,ren13}. However, evolution of granular bed structure under sub-critical hydrodynamic shear has not been studied to understand the corresponding dynamics in any significant detail. }   

Here, we use internal imaging and results drawn from the study of granular compaction and jamming to understand the response of a bed composed of monodisperse spherical grains to fluid flow below the threshold for bedload transport. Starting from a freshly sedimented bed, we investigate the spatial evolution of the bed at the grain scale away from the influence of side walls by using fluids with the same refractive index as the transparent grains. We then measure the spatio-temporal evolution of the bed as a function of imposed shear stress to understand the correlation of compaction, creep and the surface roughness over long times.  { The bed compacts with the granular volume fraction increasing logarithmically over time, up to a value associated with the jamming transition for spheres, similar to vibrated granular systems where the mechanism by which the energy is supplied to the system is different. We show that the roughness of the bed is uncorrelated with the decrease in grain movement over time. Rather, armoring occurs as the granular bed approaches random close packing with a time scale which depends on the applied shear rate.}

\section{Experimental Apparatus}
\begin{figure}
\begin{center}
\includegraphics[width=0.8\columnwidth]{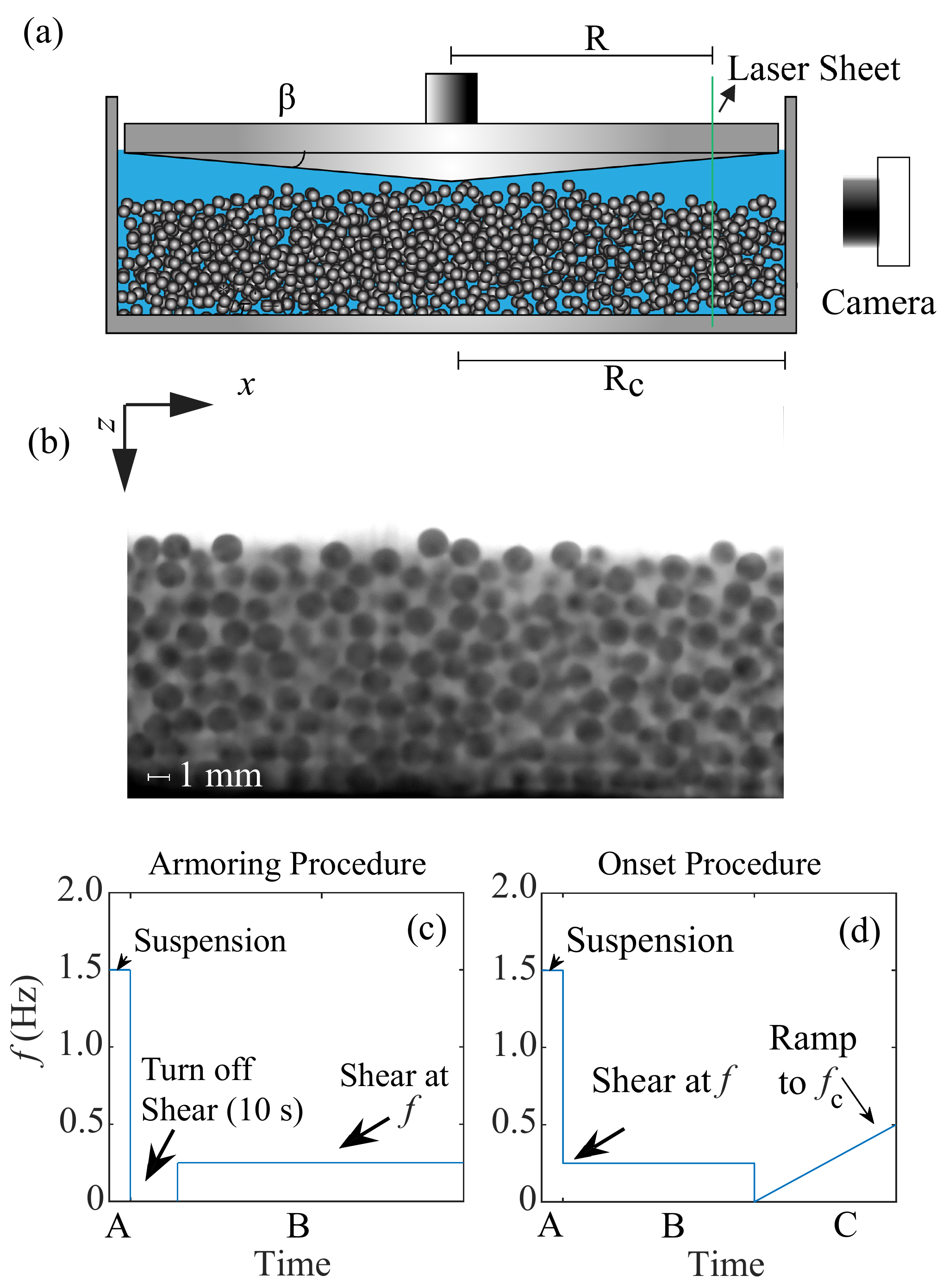}
\end{center}
\caption {(a) A schematic of the experimental apparatus which consists of a  cylindrical container filled with transparent grains and fluid.  The top plate is rotated about its axis to apply shear. { (b) An image of the grains inside the bed which appear dark in contrast with the fluid which fluoresces bright. (c) The {\it armouring} procedure used to apply shear to bed over time. The grains are initially suspended (step A), and then shear is applied by rotating the top plate at a fixed frequency (step B) after shear is turned off for 10 seconds. (d) The {\it onset} procedure used to find the threshold for eroding a bed which has been initially prepared by suspending the bed (step A) and applying a prescribed shear corresponding to $f$ for a time duration $T$ (step B). Then, the driving is turned off before being ramped up linearly to deterimine the threshold of erosion as a function of $f$ and $T$.} 
}
\label{fig:apparatus}
\end{figure}

Fig.~\ref{fig:apparatus}(a) shows the schematic of the circular flume apparatus consisting of a cylindrical container, with radius $R_c = 90$\,mm, and with a conical top plate which can be rotated with prescribed frequency $f$ to apply uniform shear to the fluid, as in Ref.~\cite{allen17}. Acrylic spheres with diameter $d = 1.52 \pm 0.01$\,mm,  density $\rho_g=1180$\,kg\,m$^{-3}$, and index of refraction $n = 1.4923 \pm 0.0003$ are filled to a height of approximately $10d$. The bottom of the container has a roughness $k = 0.5$\,mm to prevent hexagonal crystallization from occurring to model typical random packings in deep granular beds. The fluid has density $\rho_f = 1002$\,kg\,m$^{-3}$ and kinematic viscosity $\nu = 0.018$\,Pa\,s, and its refractive index is adjusted to be within 0.02\% of that of the grains using a mix of oils~\cite{siavoshi96}. In the case of a flat bed, this conical geometry results in circular fluid flow with a uniform shear rate $\dot{\gamma}$ across the bed surface given by $\dot{\gamma} = 2 \pi f /\tan{\beta}$, where $\beta$ is the angle between the conical surface and the bed surface~\cite{kudrolli16}. {} This circular flume geometry is used because it allows us to mimic a larger system than would be otherwise possible and conduct our experiments under steady state driving conditions for long times. Further, it allows us to maintain the refractive index matching properties of our fluid more easily.

\subsection{Internal visualization}

{To find the structure and motion of the bed,} we image a vertical plane inside the bed at a distance of $R=72$\,mm from the central axis which is also $12d$ away from the side walls to avoid any direct influence of the boundaries. At this relatively large $R/d \approx 50$ and a sufficiently small viewing area, the region of interest can be modeled as planar, with flow moving from the left to the right. A Cartesian coordinate system is used to describe the system with the $x$-axis along the flow, the $z$-axis along gravity, and the $y$-axis pointing into the plane. 

We use a camera combined with a high pass filter to image the system illuminated with a vertical 532\,nm laser sheet. A sample image is shown in Fig.~\ref{fig:apparatus}(b). The centroid of the dark pixels corresponding to each grain is used to determine its position within $0.02d$ in the $x-z$ plane~\cite{allen17}. Further, the diameter of the grain crosssection in the illumination plane is used to track grains within a distance $y\pm 0.22d$ from the center of the laser sheet. { In complementary experiments, we also view the bed from the top through the transparent plate by switching the position of the camera and the laser. This allows us to observe a larger surface area to find the onset of erosion with greater statistical significance as erosion occurs first at the surface before growing deeper. 

Limited experiments were also performed with glass beads which have higher relative density compared with acrylic beads, and similar overall slow relaxation over extended periods was observed. However, because we had access to larger number of beads and thus greater bed heights, we focus on experiments with acrylic beads in this paper.}

{  
\subsection{Shear protocols}

Two systematic experimental procedures are used in our studies as sketched in Fig.~\ref{fig:apparatus}(c,d) to examine the evolution of the bed as a function of applied conditions. In the {\it armoring} procedure, which is mainly used in our study, the bed is fully suspended by rotating the top plate with a frequency $f = 1.5$\,Hz, and then setting $f =0$\,Hz to turn off shear for 10\,seconds. The bed sediments and appears essentially stationary to visual inspection over this time. The sedimentation time scale of the grain in the fluid to fall through its diameter is given by $$t_s= \frac{18 \nu}{(\rho_p-\rho_f)gd}.$$ Substituting in the material properties, we find $t_s = 0.11$s. Thus, the chosen 10\,s corresponds to approximately $100t_s$ which is apparently sufficient for all the grains to fall out of suspension. The top plate is then rotated at a prescribed constant frequency to study the bed evolution over time. This procedure is used to identify the frequency $f_c$ for onset of steady state erosion as well as the armoring of the bed as we will discuss further in the paper. 

In the second {\it onset} procedure used to determine the onset of erosion depending on applied shear history, the top plate is rotated at a high frequency as in the other procedure. Then, the rotation frequency $f$ is reduced to one corresponding to a particular sub-critical shear for a prescribed wait time $T$. Shear is turned off and $f$ is then ramped up linearly to $f_c$ to identify the onset of erosion as a function of prescribed sub-critical shear or wait time $T$ with complementary visualization measurements of the surface in a $22d \times 25d$ window.

To identify $f_c$, we initially surveyed the bed over an hour to visually examine if grains continue to roll over the surface to determine the initial range of shear. Once, we narrowed the range of applied frequency, we then examined the bed after 24 hours  using the {\textsl armoring} procedure.  After this time, the surface is imaged over a 10 minute interval, and we determine if a particle rolls out of its pocket over this duration. The lowest frequency where no motion is observed using this protocol is then identified as $f_c$. This was found in~\cite{allen17} to be robust to ramping from above or below as well. This critical frequency is used to determine the stress discussed next. 
}

{

\subsection{Measurement of applied shear stress}
\label{app:shear}
A torque sensor attached to the container is used to measure the shear stress using a torque sensor as in our previous experiments~\cite{allen17}. The bed is sheared and steady state driving conditions are identified when the required torque corresponding to a prescribed rotation rate is unchanged. In the case of a flat bed, the strain rate $\dot{\gamma}$ in the fluid can be assumed to be constant at all points on the bed and given by $\dot{\gamma} = 2 \pi f/\tan \beta$, where $\beta = 5^0$ corresponding to angle which the bed makes with the cone surface. 
Then, we can relate the measured torque to the constant shear stress acting on the bed which equals the shear stress $\tau$ applied by the fluid under steady state conditions. Hence, the measured torque $N = \int_0^{R_c} dr 2 \pi r^2 \tau $, where $r$ is the radial coordinate from the center of the container, and the limits of the integral correspond to the container center and its radius $R_c$. 
For uniform $\tau$ across the bed surface, we then have
$$\tau= \frac{3 N}{2\pi R_c^3}$$ from the measured torque $N$. 

We plot $\tau$ as a function of strain rate $\dot{\gamma}$ in Fig.~\ref{fig:torque}. The shear stress acting on a planar flat surface is also measured by substituting the granular surface with a flat plate and the measured values are also plotted for reference in Fig.~\ref{fig:torque}. One observes that the shear stress in both cases increases essentially linearly as a function of the shear rate up to the onset of bedload transport. This linear increase confirms that the fluid flow in the system is in the linear viscous regime. From these measurements and direct observations, we find the critical shear stress required to dislodge grains and obtain steady-state bed transport to be $\tau_c = 0.68 \pm 0.07$\,N\,m$^{-2}$.

\begin{figure}
\includegraphics[width=0.75\columnwidth]{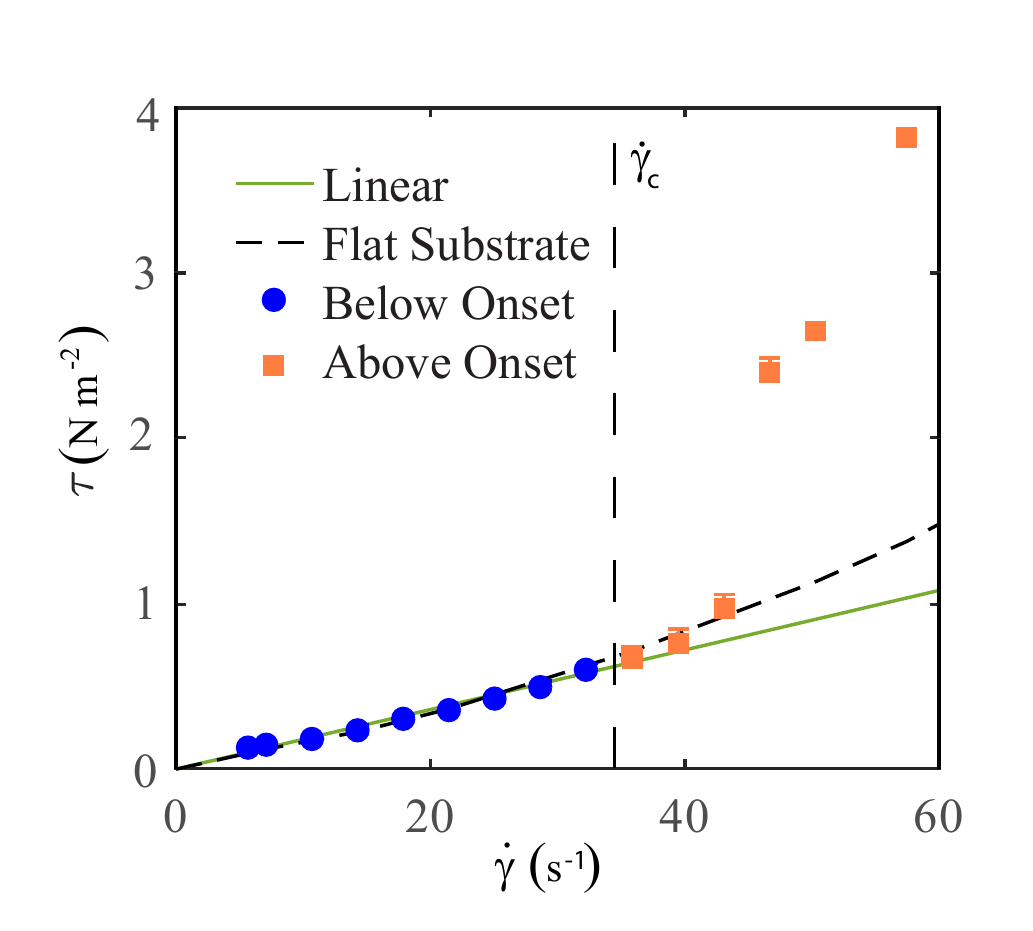}
\caption {{ The measured applied shear stress $\tau$ normalized by the critical shear stress $\tau_c$ at the onset of bedload transport as a function of strain rate $\dot{\gamma}$. $\tau/\tau_c$ is observed to increase linearly before the onset of bed load transport. The shear stress corresponding to a flat bottom (solid) surface is also plotted for reference.}  
}
\label{fig:torque}
\end{figure}

Above $\tau_c$, we observe from Fig.~\ref{fig:torque} that shear stress corresponding to the fluid increases somewhat nonlinearly. This indicates that inertial effects start to become important at these shear rates. Nonetheless, $\tau$ increases even more rapidly in the granular system with $\dot{\gamma}$, due to an increase in the effective viscosity of the fluid, because of the increasing suspension of the grains as discussed in our previous study~\cite{allen17}. In this study, we are focused on the $\tau/\tau_c < 1$ regime, and thus the flow can be assumed to be in the viscous regime based on the linear increase in $\tau$ with shear rate over that parameter range.  

In order to nondimensionalize the applied stress, and to compare to the literature on erosion, we use the normalized shear stress $\tau^*$, also called the Shield's number. This parameter is used to characterize the onset of erosion and is obtained by normalizing $\tau$ by the gravitational stress, due to the top layer,  $(\rho_g -\rho_f)gd$ and is thus
\begin{equation}
\tau^* = \frac{\tau}{(\rho_g -\rho_f)gd}.
\label{eqn:torque}
\end{equation}
Therefore, the normalized critical stress and Shields number is $\tau^*_c =  0.19\pm 0.02$. This value is consistent with previous value reported in the viscous flow regime~\cite{hong15},  and the Shields's curve~\cite{buffington97}, but higher than the values reported earlier in a slightly different circular flume with a rectangular cross section~\cite{charru04}. 

\section{Armoring}
\begin{figure}
\includegraphics[width=.95\columnwidth]{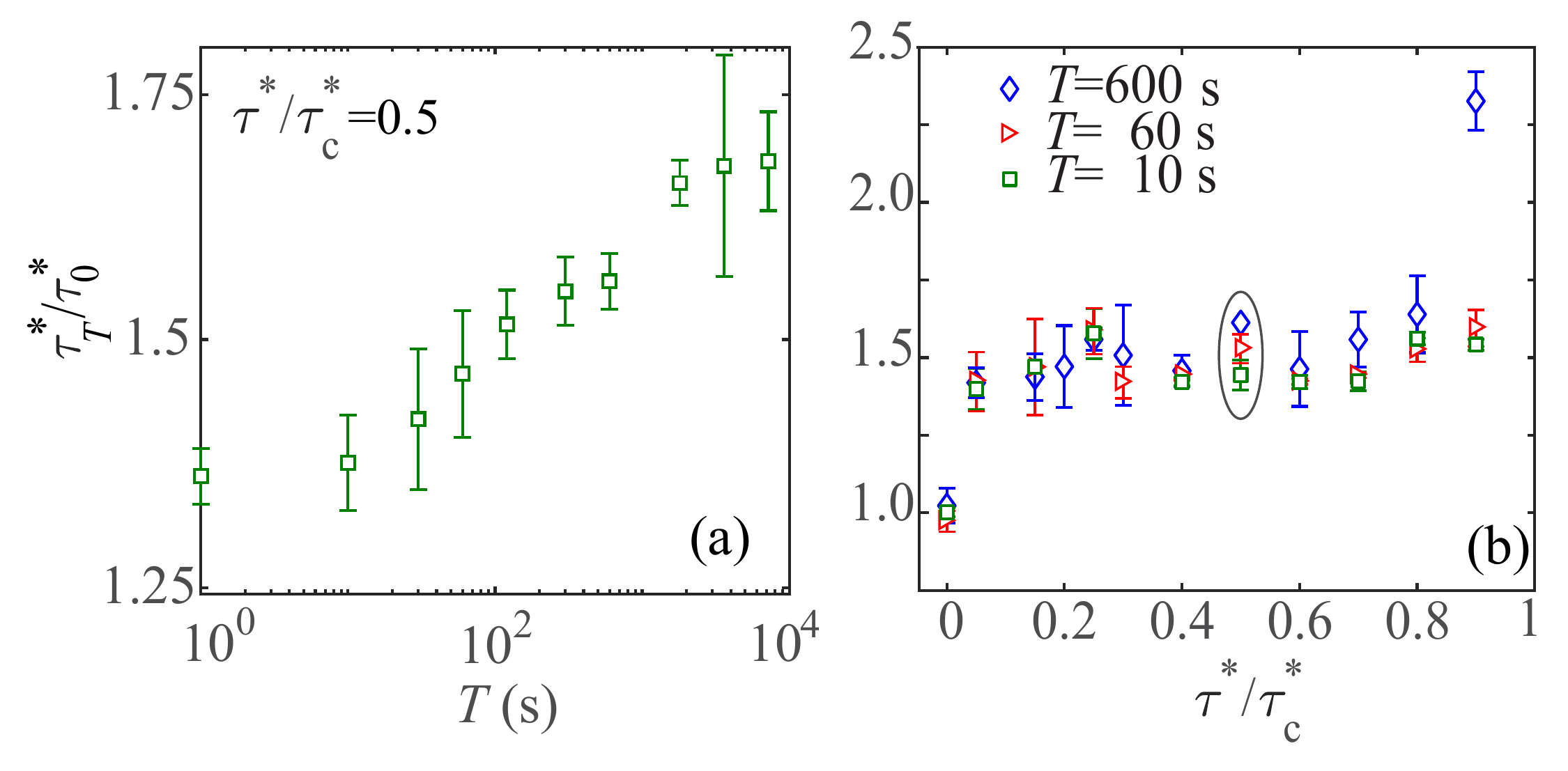}
\caption{{ (a) Measurement of the onset of transient erosion due to preshearing at half the steady state shear $\tau^*_c$ needed to erode particles at times ranging from seconds to hours. (b) Transient erosion onset for the first 10 minutes with frequencies, stresses, that approach the onset of steady state shear.  The error bars represent the rms variation of at least 10 measurements in both graphs. The data points circled by the dashed line corresponds to the stress reported in (a).
}}
\label{fig:distime}
\end{figure}

We first demonstrate that the bed strengthens when it is sheared first under sub-critical driving conditions using the {\it onset} procedure. Difference imaging obtained by subtracting images separated by 0.5 seconds is used to detect grain movement. We further ensure that grains continue to move at later times to identify onset unambiguously because bursts of isolated erosion can occur at slightly lower values. A criteria where about 1\% of the grains at the surface start to move is then used to decide onset.  Practically this is found to be the most robust method we found to identify onset. Difference images can also uncover grains that shift or roll in local pockets at earlier time which corresponds to lower applied shear rate due to the linear ramp rate. Similar overall trends are observed but the variance is found to be much higher if this is used as a criteria to identify onset of erosion.

Fig.~\ref{fig:distime}(a) shows that the recorded normalized shear stress $\tau^*_T$ at which the bed is observed to erode as a function of the preshear time $T$. Here, we have normalized the observed value with respect to the value $\tau^*_0$ observed when no shear history is present, i.e. $T=0$\,s.  We observe that $\tau^*_T$ increases systematically as $T$ is increased from 10 second to 2 hours.   
Fig.~\ref{fig:distime}(b) further shows the effect of changing applied shear on observed $\tau^*_T$ for three different $T$. A significant increase in $\tau^*_T$ is found as soon as the any amount of sub-critical shear is applied.  Further, the relative strengthening is not very sensitive to the wait time $T$ except as the critical shear rate to observe steady state bedload transport is approached. Thus, the strength of the shear, and time over which it is applied, is observed to lead significant strengthening and armoring of the bed consistent with previous observations~\cite{charru04,hong15}.   

Having demonstrated the considerable impact of the shear history of the bed on the observed onset of bed erosion, we next examine the effect of the shear on the bed structure.
}

\section{Evolution of bed structure}

\subsection{Rearrangements with depth}

\begin{figure*}
\includegraphics[width=0.65\textwidth]{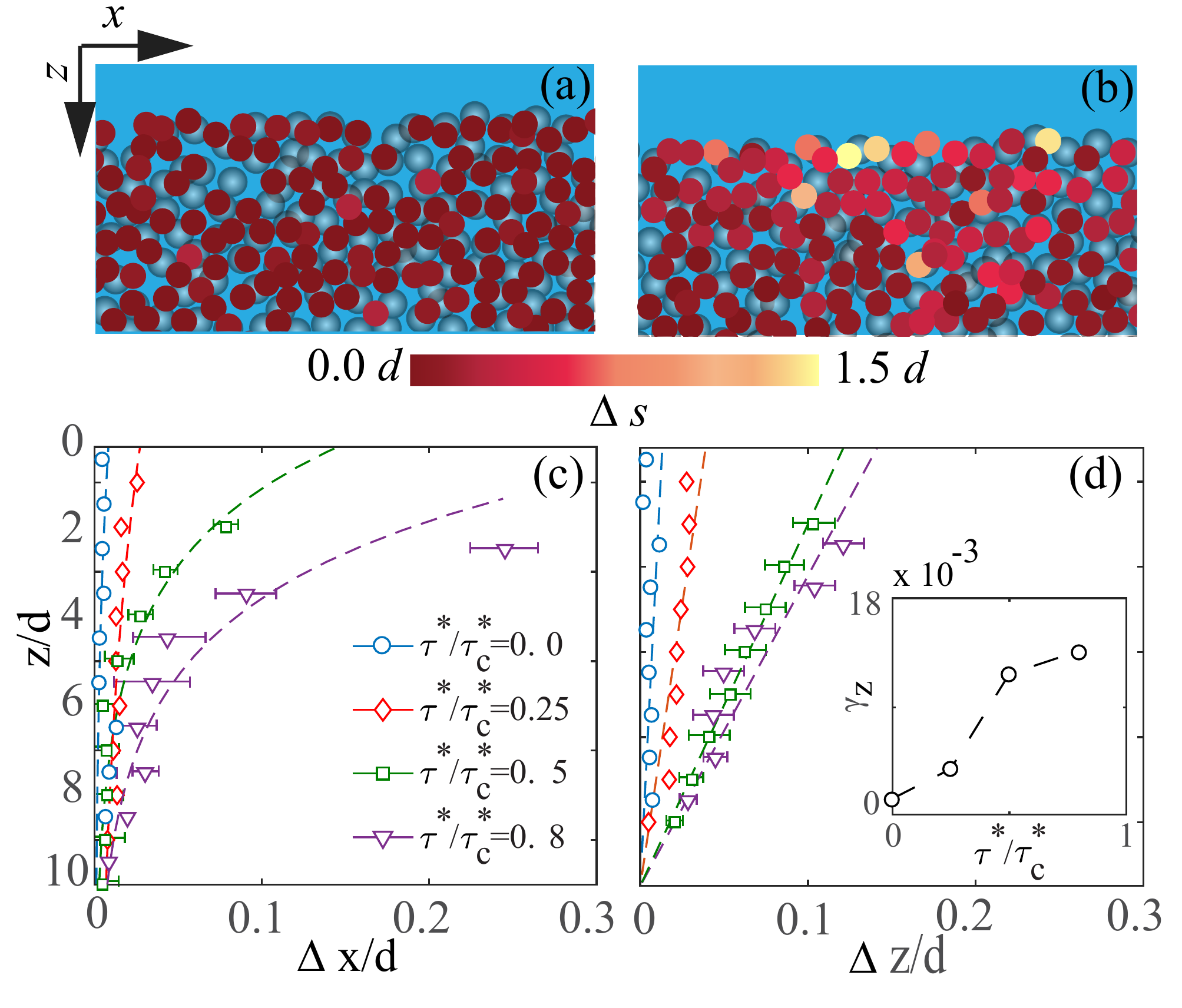}
\caption{(a) The movement of particles in the bed in the first 90 minutes of preshear at no shear stress $\tau^*/\tau^*_c=0.0$ and (b) at  $\tau^*/\tau^*_c=0.8$ where the color goes from dark to light with increasing particle movement.  We see that particles move even at no shear but there is greater movement at higher shear.  Looking more closely in a short segment where we can track all the particles $t=30-90$ seconds (c,d) are the displacement of particles as a function of distance from the surface in flow $x$ and gravity $z$ directions respectively.  We see an exponential behavior in the flow movement while the bed shifts down linearly with depth compacting uniformly.
}\label{fig:motion}
\end{figure*}

Fig.~\ref{fig:motion}(a) and Fig.~\ref{fig:motion}(b) show examples of grain positions in a vertical slice in the bed recorded through 90 minutes, corresponding to shear stresses of $\tau^*/\tau^*_c = 0.0$ and $0.8$ respectively. The data is obtained using the {\it armoring} procedure sketched out in Fig.~\ref{fig:apparatus}(d) in order to have well defined shear history conditions. Here, grains which remain within $y = \pm 0.16d$ are tracked and analyzed.  The magnitude of displacement of the individual grains $s$ in the plane over this time interval is denoted using the color map to capture the bed evolution. One observes that both example rearrange, including the one with no shear, with greater motion occurring for $\tau^*/\tau^*_c = 0.8$.  

We examine the displacements inside the bed by observing the motion of the grains over a time $t= 30$ and 90\,s. The grain displacements in the same flow and gravity directions are plotted as a function of depth $z$ in Fig.~\ref{fig:distime}(c,d), respectively.  
One observes that the bed creeps forward faster and settles further with increasing shear stress. Moreover, the creep along the flow direction appears to decay somewhat exponentially with depth as shown by the fits in Fig.~\ref{fig:distime}(c). The decay length from the exponential fit in the case of the higher shear rates, where a meaningful variation occurs, is found to be $2.5d \pm 0.1d$. This decay is similar to the length scale over which grain speeds exponentially decay into the bed for $\tau^* > \tau^*_c$~\cite{allen17}, and  was observed to be common to dry granular beds in gravity which are sheared horizontally at the top~\cite{siavoshi96,hennan13}. 

On the other hand, the linear compaction with depth at all shear rates implies that the bed settles uniformly as grains rearrange in gravity. The strain gradient $\gamma_z = -\Delta z/z$ obtained from the linear fit is shown in Fig.~\ref{fig:distime}(d) and observed to be non-zero and increase significantly with $\tau^*$. { Such a linear increase would imply that the volume fraction of the bed increases uniformly into the bed, an issue we will examine more closely later in the discussion. But, we focus first on the surface because the greatest displacements occur near the surface  as observed in Fig.~\ref{fig:motion}. }

\subsection{Surface creep}
We now examine the motion of the bed surface to gauge its creep  over long times by focusing on the grains between $0 < z < 2d$, over 48 hours under steady $\dot{\gamma}$ conditions. The average normalized displacement in the flow direction $s_x = \langle \Delta x \rangle $ and the gravitational direction $s_z = \langle \Delta z \rangle$ is plotted in Fig.~\ref{fig:distime}(a) and (b), respectively. Time $t$ is normalized by the sedimentation time scale in our system $t_s$ of a grain in the fluid due to gravity, i.e. $t_s = 18 \nu/(\rho_p-\rho_f)gd = 0.11$\,s, which is a typical time scale relevant to understanding grain-fluid systems. We observe that the grains continue to creep over the entire duration of the experiments over a million times $t_s$. Although, the overall rates are observed to decrease systematically over time, the bed creeps nonetheless over the entire time interval studied. The bed also initially settles rapidly, before slowing over time.

\begin{figure*}
\includegraphics[width=0.9\textwidth]{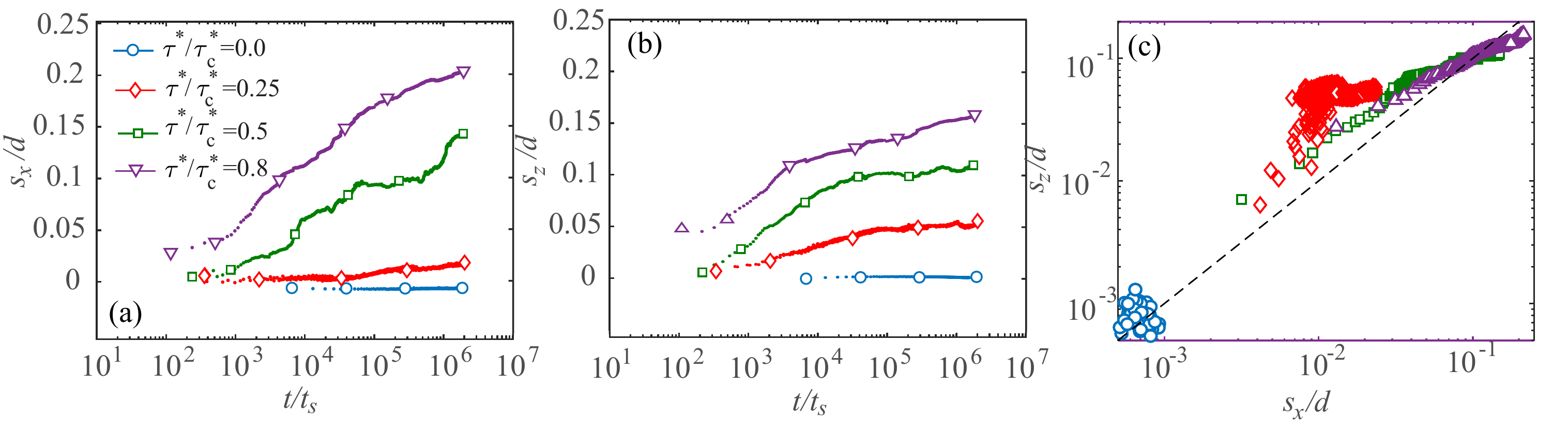}
\caption{The displacement of surface particles normalized by the particle diameter during preshear in (a) the flow direction, $x$ and (b) the direction of gravity $z$.  We see a quick initial increase but particles continue to move at long times. (c) $s_x$ versus $s_z$ is scattered for $\tau^*/\tau^*_c=0.0$. However, $s_x$ increases systematically faster compared with $s_z$ as $\tau^*/\tau^*_c$ is increased.
}\label{fig:disp}
\end{figure*}

We further examine the surface displacements of particles between $0 < z < 2d$, averaged over five different experiments up to $2\cdot 10^6 t_s$.  We measure the displacements in both the gravitational $s_z = \langle \Delta z \rangle$ and flow $s_x = \langle \Delta x \rangle $ directions Figure~\ref{fig:disp}(a) and (b), respectively over steady conditions. We observe that the grains continue to creep over the entire duration of the experiments over a million times $t_s$. Although, the overall rates are observed to decrease systematically over time, the bed creeps nonetheless over the entire time interval studied. Complementarily, the bed also initially settles rapidly, before slowing over time.

We plot $s_x$ versus $s_z$ in Fig.~\ref{fig:disp}(c) to examine the correlation of the observed bed creep and consolidation. In the case of $\tau^* = 0$, we observe a scatter of points around the line with slope one indicating that the rearrangements are somewhat uncorrelated in the horizontal and vertical direction. The amount of creep appears to be correlated with the amount of movement in the bed for $\tau^* > 0$, with greater creep corresponding to greater consolidation.    While the amount of creep is lower than the displacement along gravity at lower $\tau^*$,  the creep increases faster than the compaction due to gravity for $\tau^* \rightarrow \tau^*_c$.

\subsection{Evolution of granular volume fraction}

\begin{figure*}
\includegraphics[width=0.65\textwidth]{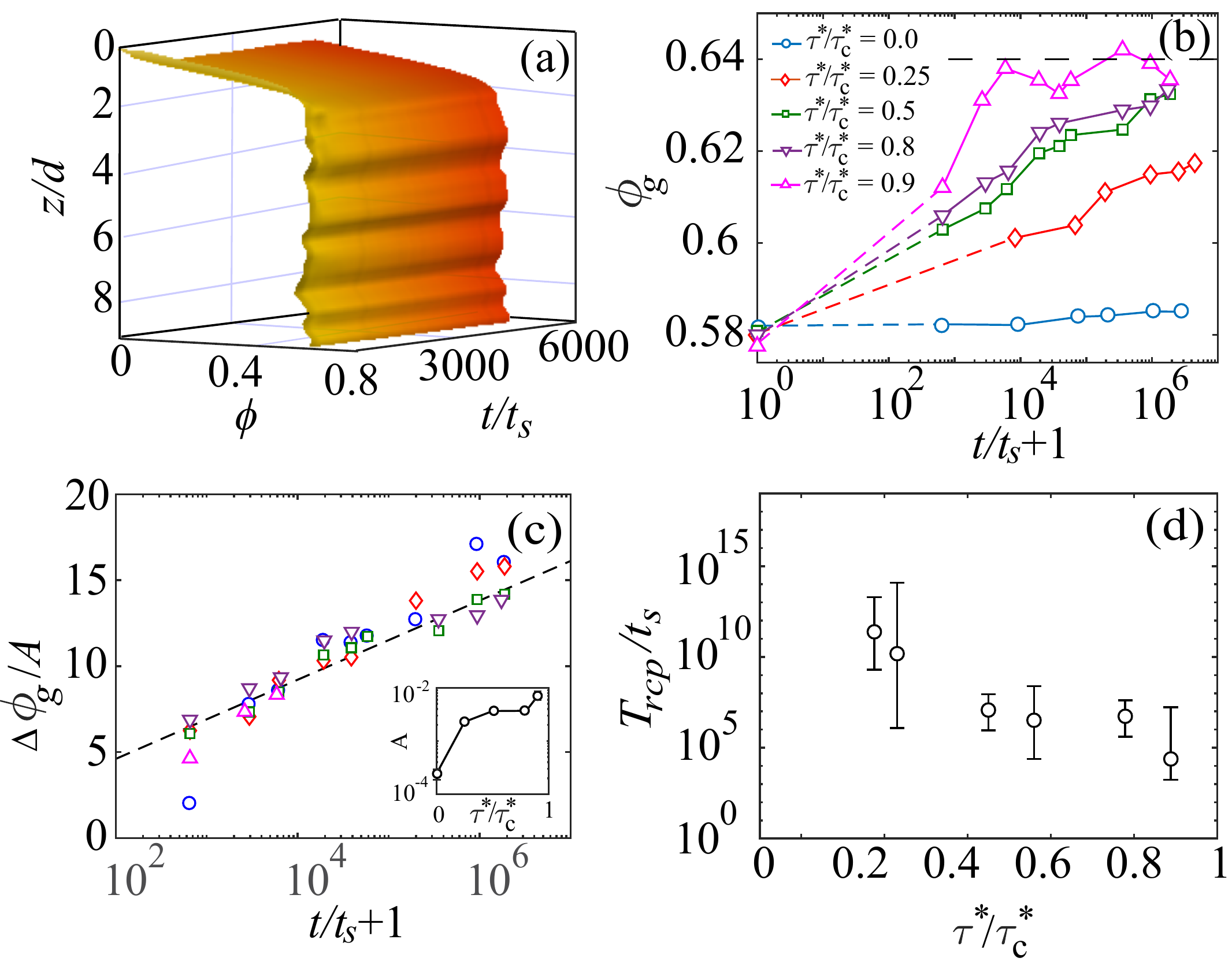}
\caption{(a) The volume fraction $\phi$ evolution as a function of depth for $\tau^*/\tau^*_c =0.8$ averaged over 10 experiments. (b) Average granular volume fraction $ \phi_g$ as a function of time $t$ rises increasingly rapidly with $\tau^*$ before approaching $\phi_{rcp}$ indicated by dashed horizontal line.  (c) The change in $\phi_g$ scaled with $A$ in the regime where $t/t_s \gg 1$ and where $\phi_g < \phi_{rcp}$. The data is observed to collapse on the line corresponding to Eq.~\ref{eq:log}. Inset: The fitted $A$ as a function of $\tau^*$. (d) The estimated time $T_{rcp}$ required to reach $\phi_{rcp}$ decreases rapidly with increasing applied sub-critical shear. 
}\label{fig:phi}
\end{figure*}

We next study the evolution of the volume fraction of the grains $\phi$, to understand these trends in the bed relaxation. Fig.~\ref{fig:phi}(a) shows an example of $\phi$ variation with depth, which is observed to increase sharply at the bed surface, reaching an essentially constant value within fluctuations for $z/d > 2$. The bed surface settles with increasing $t$ as evident from the shift downward of $\phi$ in the $z-t$ plane in Fig.~\ref{fig:phi}(a). This is consistent with the displacement of the bed surface over time seen in Fig.~\ref{fig:distime}(b), where $s_z$ was observed to increase with $\tau^*$. We also examine the average granular volume fraction in the bed $\phi_g = \langle\phi \rangle$ 
in Fig.~\ref{fig:phi}(c) to quantify the net effect of this relaxation as a function of time, where $\langle ..\rangle$ corresponds to averaging over the depth $z/d>2$. We find that $\phi_{g}$ increases over time in each case, with faster increase for higher $\tau^*$. Thus, a higher packing fraction is reached for higher $\tau^*$ over the same 48 hours time interval. We also observe in the case of the higher $\tau^*$ that $\phi_g$ rises, and saturates around $0.64$. This volume fraction corresponds to the random close packing fraction $\phi_{rcp} = 0.64 \pm 0.02$ that spherical grains approach when the glass transition is reached~\cite{berryman83,ohern02}.  

It has been also shown that the volume fraction of randomly created spherical grains are well below the maximum volume fraction for spheres $\phi_{max} = 0.74$ ~\cite{scott69}, unless special protocols are used~\cite{apanatescu12,apanatescu14}. In fact, the typical value of $\phi$ reached in spherical granular matter is in fact well below $\phi_{rcp}$ due to friction between grains~\cite{onoda90}. When vibrated or tapped, such packings compact rapidly at first with compaction growing logarithmically over long times as they approach $\phi_{rcp}$~\cite{knight95,philippe02,ribiere05}. This logarithmic slowing dynamics has been explained using the parking lot model~\cite{bennaim1998}, where increasingly large number of grains have to rearrange collectively to create space which is sufficient to fit an additional grain and thus increase the total volume fraction. While previous studies on vibrated granular materials have shown similar dynamics in terms of the total height of a vibrated granular column and capacitive measurements~\cite{knight95,ribiere05,philippe02,richard05}, our study is the first to discuss the spatial evolution inside the bed starting with individual grain rearrangements as shown in Fig.~\ref{fig:motion}. 

To compare the compaction dynamics, we evaluate the change of volume fraction from the initial value before shear is applied
\begin{equation}
\Delta \phi_g = \phi_g - \phi_o = A \ln (1 + t),
\label{eq:log}
\end{equation} 
where, $\phi_o$ and $A$ are constants related to initial conditions and system properties, and time has been shifted by one to avoid the singularity at $t=0$. Fig.~\ref{fig:phi}(c) shows $\Delta \phi_g$ obtained in our experiments scaled by $A$ plotted versus time $t$ corresponding to Eq.~\ref{eq:log}, with the fitting constant $A$ shown in the inset. We observe a good collapse of the data over 4 orders of magnitude in the regime after the initial rapid transient regime and before $\phi_{rcp}$ is reached. Thus, we observe similar granular compaction dynamics under sub-critical hydrodynamic shear as in vibrated systems~\cite{knight95,ribiere05,bennaim1998,philippe02,richard05}. Like the acceleration strength in the vibration experiments, the compaction logarithmically approaches $\phi_{rcp}$ faster as $\tau^*$ is increased.

{ In vibrated systems, the granular bed has energy imparted to it by external taps or vibration~\cite{knight95} which show a similar logarithmic growth of packing fraction towards $\phi_{rcp}$. In the sheared system, similar slow down occurs as the bed volume fraction increases toward the random close packing limit, as greater and greater number of grains have to move cooperatively for grains to settle further. This becomes less likely with time, and has been argued to be the basis of the logarithmic behavior in time in vibrated systems~\cite{bennaim1998,richard05}.  Unlike vibrated or tapped systems, shear stress is continuously applied by the fluid in our system. However, grain rearrangements which are discrete in time also occur due to granular collisions as grains are dislodged at the surface that can percolate momentum into the bed. Further, theoretical work is needed to understand the origin of this similar dynamics in these systems.}

In Fig.~\ref{fig:phi}(d) the time $T_{rcp}$ that it takes the bed density to reach random close packing is shown assuming the logarithmic increase continues until the bed saturates at $\phi_{rcp}$ { to further characterize the system.  Here, the time is extrapolated from the fits to Eq.~\ref{eq:log}, and only in the case of $\tau^*/\tau^*_c = 0.9$ do we actually see a saturation in $\phi$ around $\phi_{rcp}$, with some fluctuations in our experiments.} One finds that the time that it takes for the bed to compact changes dramatically with $\tau^*$. Although some compaction is observed at $\tau^* \approx 0$, the time scales obtained to reach $\phi_{rcp}$ are greater than the lifetime of the universe, and therefore not shown. At somewhat larger $\tau^*$, we find that $T$ decreases from about six centuries to the highest $\tau^*/\tau^*_c$ studied, where $T$ is of order of half an hour as the critical shear stress $\tau^*_c$ is approached.

\begin{figure}
\includegraphics[width=0.7\columnwidth]{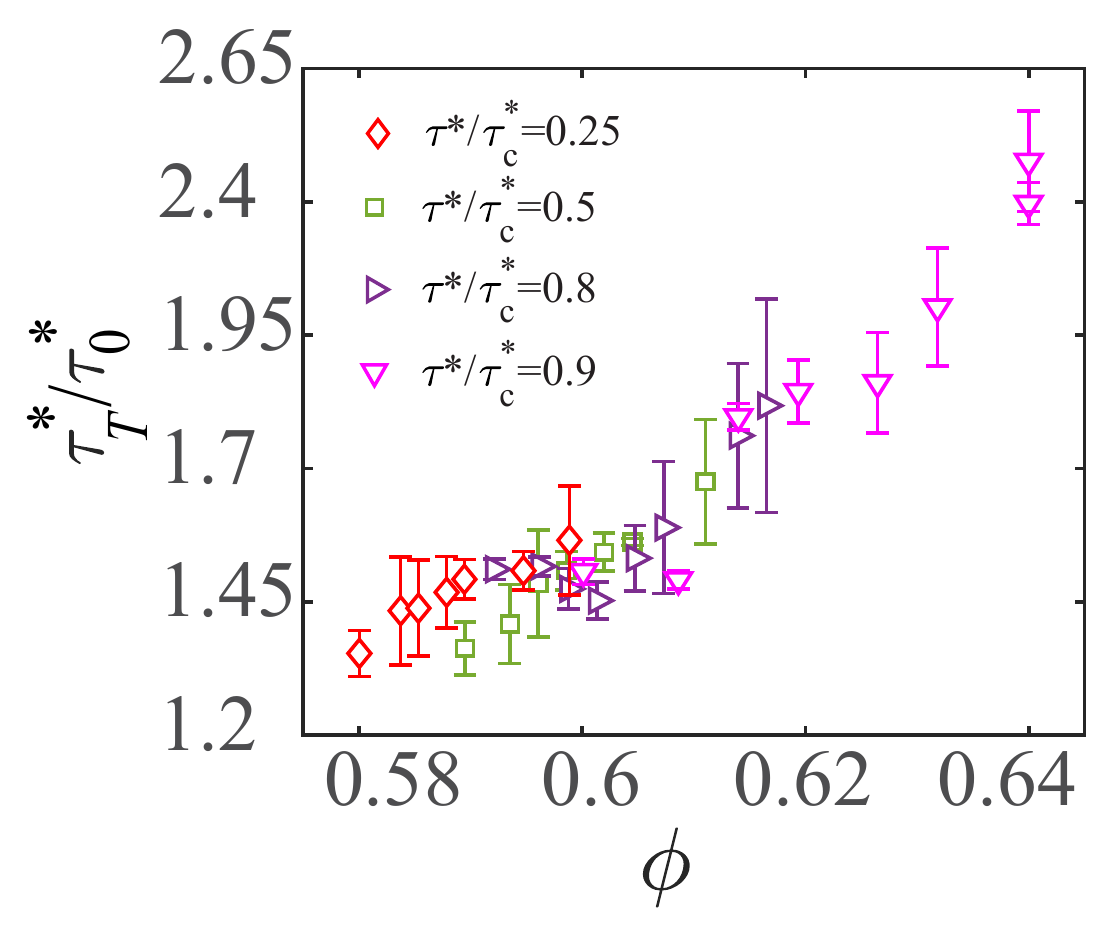}
\caption{{ The observed shear stress at onset is observed to increase systematically with the granular packing fraction in the bed $\phi$ regardless of preshearing strength. Error bars indicated correspond to at least 10 experimental runs for each condition.}  
}\label{fig:armor}
\end{figure}

{
To examine the effect of granular volume fraction on the observed shear stress required to erode the bed, we plot $\tau^*_T$ obtained as a function of $\phi$  in Fig.\ref{fig:armor}.  We find that there is indeed a systematic increase in the erosion threshold as the packing fraction increases. Further, the data corresponding to the various sub-critical applied  shear $\tau^*$ more or less increase together. Thus, the control parameter which determines the erosion threshold according to this plot is the granular volume fraction of the bed reached at that time, rather then the time duration or strength of the pre-shear.}

\subsection{Surface roughness}

\begin{figure}
\includegraphics[width=0.8\columnwidth]{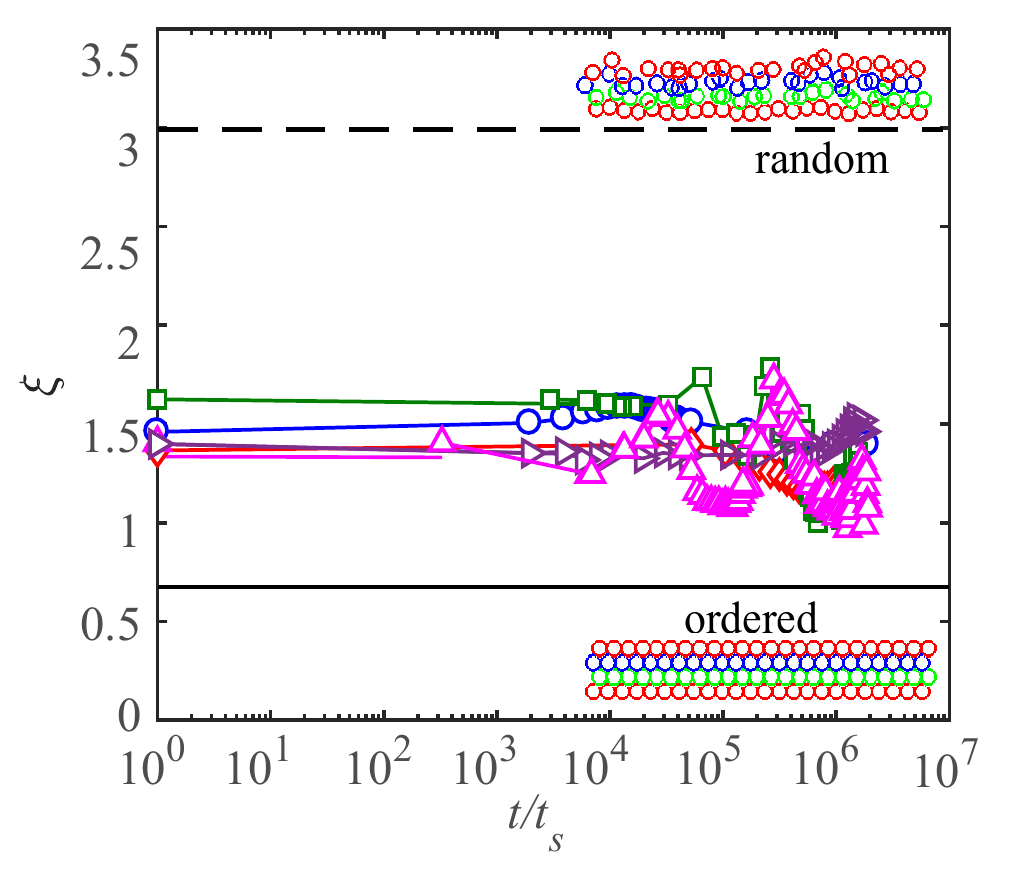}
\caption{Surface roughness $\xi$ evolution for various $\tau^*$ shown in Fig.~\ref{fig:phi}(b). The values corresponding to random and ordered hexagonal packings near the surface are also shown.  We see that the surface does not change systematically in either direction while we saw earlier that shearing for long times changes the onset of erosion.
}\label{fig:surface}
\end{figure}

We also examine the bed roughness to understand its contribution to the evolution of the creep as a function of applied stress. We obtain the roughness measure $\xi$ by calculating the change of depth $z$ where $\phi$ increases from $0.1$ to $0.4$, corresponding to roughly half the observed variation in $\phi$. Plotting this measure $\xi$ as a function of time in Fig.~\ref{fig:surface}, we find that the roughness of the bed can vary somewhat, but not systematically, over the entire duration of the experiment in the case of each $\tau^*$. 

For reference, we also calculated $\xi$ corresponding to the case where the bed was ordered in a triangular lattice and a random surface.  The random surface is created by randomly moving grains from a triangular lattice using a uniform distribution between 0 and $d$ resulting in the grain positions shown in Fig.~\ref{fig:surface}. We observe that the variation in $\xi$ is relatively small compared to these two limits, and the bed surface roughness is essentially unchanged as $\phi_g$ varies from 0.58 to 0.64 over time for the various $\tau^*$ and uncorelated with decreasing rate of erosion. 

The lack of significant change toward the ordered value shows that the presence of shear does not appear to make the bed surface any smoother by selectively eroding particles that protrude further into the fluid.   Thus, we find that decrease in creep and consolidation show no significant correlation  with the observed fluctuations in bed surface roughness.

\section{Conclusions}

{ In conclusion, we show that a freshly sedimented bed consolidates and creeps slowly over long times under sub-critical hydrodynamic shear conditions.  We also clearly demonstrate that the threshold for initial erosion of a granular bed is systematically greater in case of a bed which has been sheared under sub-critical conditions. Both the duration and strength of the applied sub-critical shear is observed to have a signicant impact on the so called armouring of the bed. The internal visualization of the granular phase using refractive index matching technique allows us to obtain the spatial distribution of the observed rearrangements and uniform evolution of volume fraction. 

While a small degree of consolidation is observed when shear is absent, the application of shear stress causes rearrangements to increase more rapidly, and the grains to settle to a much greater degree over the same observation time interval.  We find that logarithmic compaction dynamics is observed, similar to those in vibrated granular beds. This is interesting because the mechanism by which stress or energy is input into these athermal systems is rather different. In our case, the stress is applied continuously, whereas, energy is added as a series of discrete taps in the case of the previously studied compation experiments.  Further, unlike the vibrated compaction experiments, we observe creep along the direction of flow because the sheared system is unbounded, and creep increases with increasing strength of sub-critical shear. This creep is observed to decay exponentially into the bed similar to observations deep inside the bed above the threshold for continous bedload transport at the surface as we reported on previously~\cite{allen17}.  

We further demonstrate that the appropriate control variable for the observed transient erosion threshold is in fact the granular packing of the bed rather then the strength or the duration of the applied shear. Hence, we demonstrate that shear history has significant impact on the time evolution of granular beds subjected to constant hydrodynamic shear even below the critical Shield's number through the consolidation of the bed. 
}
Thus, our results have broad implications for granular systems which may appear stationary at short time scales but in which significant grain movement and changes in bed strength occur over long time scales as illustrated by the armoring and change in erosion threshold.


\begin{acknowledgments}
This work was supported by the National Science Foundation Grant CBET-1335928, and the US Department of Energy Office of Science, Office of Basic Energy Sciences program under DE-FG02-13ER16401. 
\end{acknowledgments}

\end{document}